\documentclass[intlimits,twoside,a4paper]{article}

\usepackage[cp1251]{inputenc}
\usepackage[eqsecnum]{cmpj3}

\usepackage{bm}
\usepackage{xcolor}

\issue{2021}{24}{1}{13002}
\doinumber{10.5488/CMP.24.13002}
\title[The effect of non-local derivative on Bose-Einstein condensation]%
{The effect of non-local derivative on Bose-Einstein condensation%

}
\author[F.E. Bouzenna, M.T. Meftah, M. Difallah]{F.E. Bouzenna\refaddr{label1}\,\footnote{e-mail: fatmahp2000@gmail.com},
        M.T. Meftah\refaddr{label2}\,\footnote{e-mail: mewalid@yahoo.com}, M. Difallah\refaddr{label3}\,\footnote{e-mail: mosdifah@yahoo.fr}}
\addresses{
\addr{label1} Physics Department and LEVRES laboratory, Faculty of Exact Sciences, University of El Oued, 39000, Algeria\\
\addr{label2} Physics Department and LRPPS laboratory, Faculty of Mathematics and Matter Sciences, University Kasdi Merbah, Ouargla 30000, Algeria\\
\addr{label3} Physics Department and LABTHOP laboratory, Faculty of Exact Sciences, University of El Oued, 39000, Algeria\\
}
\date{Received June 20, 2020}
\begin{document}

\maketitle

\begin{abstract}
In this paper, we study the effect of  non-local derivative on Bose-Einstein condensation. 
Firstly, we consider the Caputo-Fabrizio derivative of fractional order $\alpha$ to  derive the eigenvalues of non-local Schr\"{o}dinger equation for a free particle in a 3D box. 
Afterwards, we consider 3D Bose-Einstein condensation of an ideal gas with the obtained energy spectrum. 
Interestingly, in this approach the critical temperatures $T_{c}$ of condensation for  $1<\alpha<2$  are greater than the standard one.  
Furthermore, the condensation in 2D is shown to be possible.
Second and for comparison, we presented, on the basis of a spectrum established by N. Laskin, the critical transition temperature as a function of the fractional parameter $\alpha$ for a system of free bosons governed by an Hamiltonian with power law on the moment ($H\sim p^{\alpha}$). In this case, we have demonstrated that the transition temperature is greater than the standard one.
By comparing the two transition temperatures (relative to Caputo-Fabrizio and to Laskin), we have found for fixed $\alpha$ and the density $\rho$ that the transition temperature relative to Caputo-fabrizio is greater than relative to Laskin. 
%\pacs 05.30.-d, 05.45.Df, 02.60.Lj, 03.75.Nt, 68.35.Rh
\end{abstract}

\section{Introduction}
 
The fractional quantum mechanics is a new branch in physics introduced recently by Laskin \cite{laskin2000} through generalizing the Feynman path integration to L\'{e}vy formulation. In this approach, the order of the space derivative in the Schr\"odinger equation is the non-integer number $ \alpha $ instead of 2. In this case, the modified equation is the so-called  fractional Schr\"odinger equation which is a non-local equation. On the one hand, Laskin  used the Riesz fractional derivative (FD) to find the associated energy spectrum and the wavefunctions of the studied systems. On the other hand, the authors \cite{hermann} and \cite{muslih}  used the Caputo derivative that provided different solutions.          

Afterwards, the fractional quantum mechanics was applied by Alisultanov and Meilanov to study ideal quantum statistical systems with Laskin energy spectrum \cite{ali1,ali2}. 
The latter has a fractional (non-quadratic) power of the momentum.
 Moreover, using the Green's function approach, they showed that these systems can emulate real systems with physical inter-particle interactions.        
In this context, \cite{z14}  studied thermodynamic properties of 3D ideal Fermi and Bose gases, 3D independent harmonic oscillators, and black-body radiation.
Interestingly, \cite{ali2,z14} found that the critical condensation temperatures of bosons with this spectrum are larger than in the standard ideal case.
More recently, \cite{z16,r17}  extended this approach to explore d-dimensional statistical systems. 
To the best of our knowledge, no large-scale study has revealed the different effects of fractional derivative definitions on Bose-Einstein condensation. 
%-------------------------------------------------------------------------
 
%
In this work, we investigate Bose-Einstein condensation through the Caputo-Fabrizio derivative \cite{cf} and compare our findings with those presented in  \cite{z14}. Unlike Riesz  and Caputo FD, the Caputo-Fabrizio derivative is improved, so that the kernel becomes non-singular. Therefore, this newly proposed derivative was applied to various fields of physics, for example, classical mechanics \cite{losada}, heat transfer \cite{shah}, fluid mechanics \cite{saqib}, and quantum  mechanics \cite{gomez}.

Our paper is organized as follows: section \ref{s2} studies 3D ideal Bose gas with single-particle energy spectrum derived from  a fractional Schr\"{o}dinger equation for a free particle in a hard-wall box using Caputo-Fabrizio FD. In section \ref{s3}, by using Laskin spectrum for the ideal bose gas, we  presented the transition temperature as function of the fractionnal order derivative $\alpha$.
Section \ref{s4}  discusses the obtained critical temperatures compared with those found according to Laskin's and to Fabrizio-Caputo energy spectra. Conclusion is then presented in section \ref{con}.

\section{The Bose-Einstein condensation based on Caputo-Fabrizio derivative}
\label{s2}
We consider the Caputo-Fabrizio derivative in resolving the non-local Schr\"{o}dinger equation because; (i) the Caputo-Fabrizio derivative of a constant function is exactly zero,
(ii) the Caputo-Fabrizio kernel has no singularity. Physically, these two properties are very important and desirable.

The Caputo-Fabrizio derivative of the order $\gamma$ where  $0<\gamma<1$
is given as follows \cite{cf}%
\begin{eqnarray}
 ^{CF}\mathfrak{D}^{\gamma}f(x)=\frac{(2-\gamma)M(\gamma)}{2(1-\gamma)}
\int_{0}^{x}f^{\prime}(t)\rd t
\exp\left[  \frac{-\gamma}{1-\gamma}(x-t)\right],\quad  x\geqslant0,
\label{eq2.1}
\end{eqnarray}
where $M(\gamma)=2/(2-\gamma)$ is a normalization constant that depends on $\gamma$. By applying the definition (\ref{eq2.1}), the
solution of the following non-local differential equation%
\begin{equation}
^{CF}\mathfrak{D}^{\gamma}f(x)=u(x),~~~~0<\gamma<1
\end{equation}
is given by \cite{losada}
\begin{equation}
f(x)=(1-\gamma)u(x)+\gamma%
%TCIMACRO{\dint \limits_{0}^{t}}%
%BeginExpansion
\int_{0}^{x}
%EndExpansion
u(t)\mathrm{d} t+f(0). \label{f1}%
\end{equation}
Consider now the non-local Schr\"{o}dinger equation for a free particle moving in
1D box with length $L$  \cite{bouz1,bouz2}

\begin{eqnarray}
\nonumber ^{CF}\mathfrak{D}^{2\gamma}\psi(x)~=-\left(  \frac{a_{0}^{2\gamma}}{\hbar^{2\gamma}D_{2\gamma}%
}\right)  E_{2\gamma}\psi(x),~~~~x\geqslant0~~, ~~\frac{1}{2}<\gamma<1,
\end{eqnarray}
or equivalently

\begin{equation}
^{CF}\mathfrak{D}^{\alpha}\psi(x)~=-\varepsilon_{\alpha}\psi(x),~~~1<\alpha
=2\gamma<2, \label{f2}%
\end{equation}
where the operator $^{CF}\mathfrak{D}^{\alpha}$ has the same meaning as (\ref{eq2.1}) and $x=X/a_{0}$ being the reduced space coordinate, $a_{0}$ is the characteristic length, $D_{2\gamma}$ is
a generalized coefficient and $ \varepsilon
_{\alpha}=\varepsilon
_{2\gamma}=a_{0}^{2\gamma}E_{2\gamma}/\hbar^{2\gamma}D_{2\gamma}$ is the reduced energy.
To seek for a solution of Schr\"{o}dinger equation (\ref{f2}), we take: $^{CF}%
\mathfrak{D}^{\gamma}f(x)=u(x)~$ and $^{~CF}\mathfrak{D}^{2\gamma}f(x)=^{CF}\mathfrak{D}^{\gamma
}u(x)=g(x)$. Hence, according to the solution (\ref{f1}) we obtain
\begin{equation}
u(x)=(1-\gamma)g(x)+\gamma%
%TCIMACRO{\dint \limits_{0}^{t}}%
%BeginExpansion
\int_{0}^{x}
%EndExpansion
g(t) \mathrm{d} t+u(0), \label{f3}%
\end{equation}
and
\begin{equation}
f(x)=(1-\gamma)u(x)+\gamma%
%TCIMACRO{\dint \limits_{0}^{t}}%
%BeginExpansion
\int_{0}^{x}
%EndExpansion
u(t) \mathrm{d} t+f(0). \label{f5}%
\end{equation}
Plugging (\ref{f3}) into (\ref{f5}), the solution
of  non-local differential equation $^{~CF}\mathfrak{D}^{2\gamma}f(x)=g(x)$ reads%

\begin{eqnarray}
 f(x)=(1-\gamma)^{2}g(x)+2\gamma(1-\gamma)\int_{0}^{x}g(t) \mathrm{d} t+\gamma^{2}\int_{0}^{x}\mathrm{d} t\int_{0}^{t}g(t)\mathrm{d}t 
+(\gamma x-\gamma+1)u(0)+f(0).
\end{eqnarray}

Let us now consider $f(x)=\psi(x)$ and $g(x)=-\varepsilon_{\alpha}\psi(x)$, then we derive twice from the above equation, and we find:

\begin{eqnarray}
\psi^{\prime\prime}(x)+\frac{2\alpha(2-\alpha)\varepsilon_{\alpha}%
}{4+(2-\alpha)^{2}\varepsilon_{\alpha}}\psi^{\prime}(x)+\frac{\alpha
^{2}\varepsilon_{\alpha}}{4+(2-\alpha)^{2}\varepsilon_{\alpha}}\psi(x)=0, ~~~~ \alpha= 2\gamma,
\end{eqnarray}
or%
\begin{equation}
\psi^{\prime\prime}(x)+2\Lambda\psi^{\prime}(x)+\omega_{0}^{2}\psi(x)=0,\label{f40}
\end{equation}
where $\Lambda=\alpha(2-\alpha)\varepsilon_{\alpha}/\left( 4+(2-\alpha
)^{2}\varepsilon_{\alpha}\right) $ and    $\omega_{0}=\alpha \left[ \varepsilon
_{\alpha}/\left( 4+(2-\alpha)^{2}\varepsilon_{\alpha}\right) \right] ^{1/2}$. The solution of equation (\ref{f40}) has the following form
\begin{equation}
\psi(x)=\phi(x) \mathrm{e}^{-\Lambda x},
\end{equation}
where $\phi(x)$ fulfills the equation 
\begin{equation}
\phi(x)+k^{2}\phi(x)=0,
\end{equation}
with $k=\left(  \omega_{0}^{2}-\Lambda^{2}\right)  ^{1/2}$ being the reduced wavenumber ($\omega_{0}>\Lambda $). Then, the solution of (\ref{f40}) is  given as follows:
\begin{equation}
\psi(x)=\mathrm{e}^{-\Lambda x}[A \sin\left(kx\right)+B \cos\left(kx\right)].
\end{equation}
By using the hard-wall boundary conditions%

\begin{equation}
\left\{
\begin{array}
[c]{c}%
\psi(0)=0\\
\psi(a)=0~
\end{array}
\right.
\end{equation}
where $a=L/a_{0}$, we find%
\begin{equation}
\psi(x)=A \mathrm{e}^{-\Lambda x}\sin(kx),\label{222}
\end{equation}
and
\begin{equation}
k=\frac{\piup}{a}n,~~~~n\in Z^{+},%
\end{equation}
here, the magnitude $ A $ is determined by the normalization condition $ \int_{0}^{a}|\psi(x)|^{2} \mathrm{d} x=1 $. 
In addition, we have

\begin{equation}
k=\frac{2 \alpha\varepsilon_{\alpha}^{1/2}}{4+(2-\alpha)^{2}\varepsilon
_{\alpha}},%
\end{equation}
therefore,

\begin{equation}
k(2-\alpha)^{2}\varepsilon_{\alpha}-2\alpha\sqrt
{\varepsilon_{\alpha}}+4 k=0.
\end{equation}
This equation has two real solutions, $\epsilon_{\alpha}^{1}$ and $\epsilon_{\alpha}^{2}$, only if
\begin{equation}
0<k<\frac{\alpha}{2(2-\alpha)}=B(\alpha).\label{f41}
\end{equation}
The first one is
\begin{equation}
\epsilon_{\alpha}^{1} =\frac{\alpha-\left[  \alpha^{2}
-4k^{2}(2-\alpha)^{2}\right]  ^{1/2}}{k(2-\alpha)^{2}}
=\alpha\frac{1-\left[  1-4k^{2}\left(  \frac{2-\alpha}{\alpha}\right)
^{2}\right]  ^{1/2}}{k(2-\alpha)^{2}}. \label{f9}%
\end{equation}
In the limit $\alpha\rightarrow2$, the solution (\ref{f9}) tends to $k$, and we recover the standard energy spectrum

\begin{equation}
E=\frac{\hbar^{2}D_{2}}{a_{0}^{2}}k^{2}=\frac{\piup^{2}\hbar^{2}}{2mL^{2}}%
n^{2},~~~D_{2}=\left(  2m\right)  ^{-1}.%
\end{equation}
The second solution is not valid because in the limit $\alpha\rightarrow2$ it does not
yield $\sqrt{\varepsilon}=k$. The energy spectrum of the equation (\ref{f2}) can be written as
\begin{equation}
\varepsilon_{\alpha}=\alpha^{2}\frac{\left[  1-\sqrt{1-4k^{2}\left(
\frac{2-\alpha}{\alpha}\right)  ^{2}}\right]  ^{2}}{(2-\alpha)^{4}k^{2}},%
\end{equation}
or
\begin{equation}
E_{\alpha}=\frac{\hbar^{\alpha}D_{\alpha}}{a_{0}^{\alpha}}\frac{\alpha^{2}}{(2-\alpha)^{4}}\frac{\left[  1-\sqrt{1-4k^{2}\left(
\frac{2-\alpha}{\alpha}\right)  ^{2}}\right]  ^{2}}{k^{2}}.\label{100}%
\end{equation}

\begin{figure}[!t]
\begin{center}
\includegraphics[width=9cm]{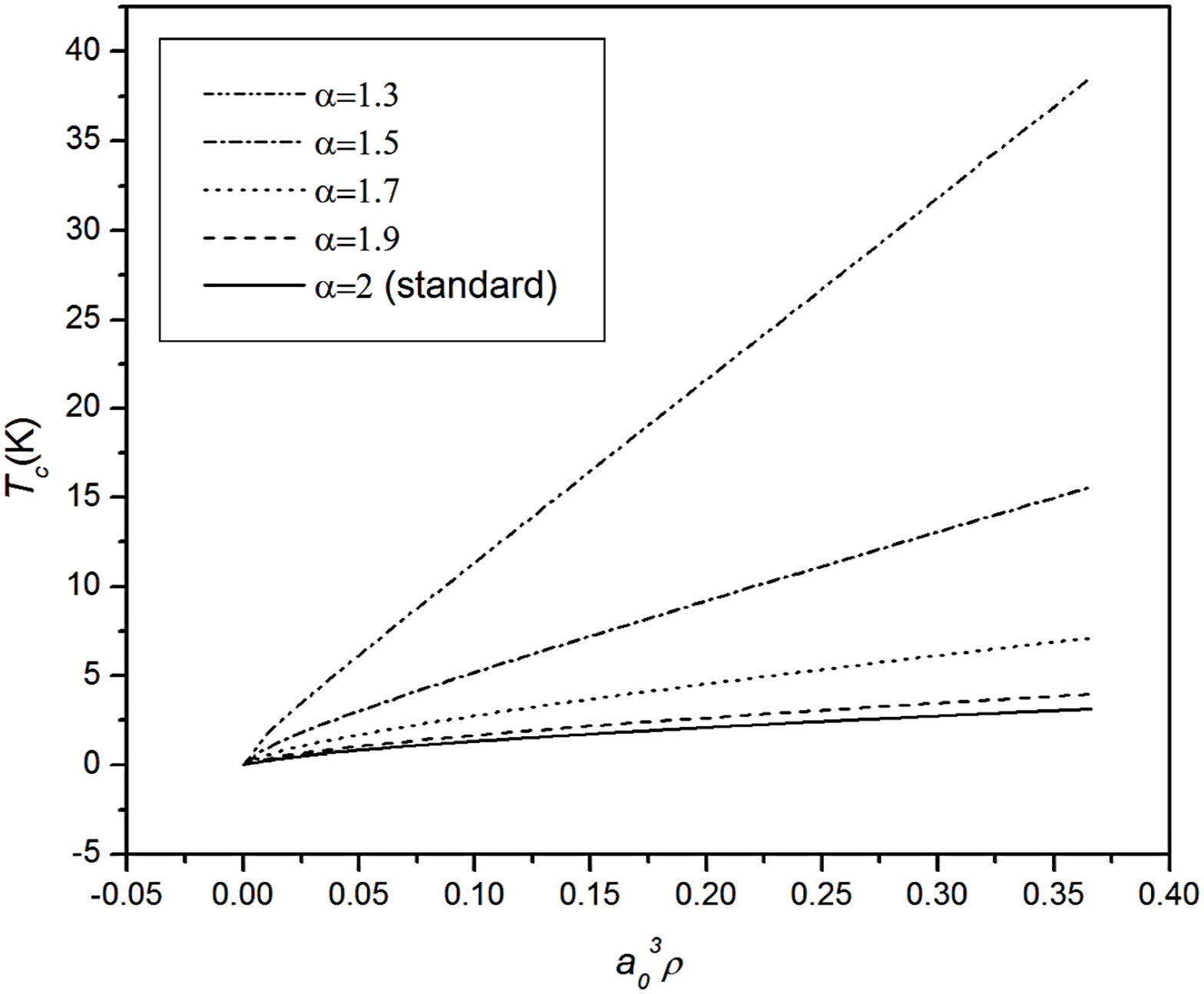}
\caption{The critical temperature curves at different $\alpha$ values within the Caputo-Fabrizio approach.}
\label{fig.1}
\end{center}
\end{figure}

From the above results we note: (i) the number of energy levels is limited by the condition (\ref{f41}), and reaches infinity at $ \alpha=2 $ [$ B(2)=\infty $, see equation (\ref{f9})]. (ii) For $ k=0 $, the energies $ E_{\alpha}=0 $ for all $ \alpha $ values.
This means that, regardless of the form of the energy spectrum, the effect of Caputo-Fabrizio definition seems to  limit the number of energy levels and to relate it to the derivative order $ \alpha $. Therefore, according to Caputo-Fabrizio approach, we expect that the critical temperature of Bose-Einstein condensation will be higher (see figure~\ref{fig.1}).
To check this result, we consider an ideal Bose gas in a hard-wall box with the volume $V=[0,L]^{3}$, so that the single-particle energy spectrum is defined by~(\ref{100}) as
\begin{equation}
 E_{\alpha}=E_{\alpha}^{1}+E_{\alpha}^{2}+E_{\alpha}^{3}
=\frac{\hbar^{\alpha}D_{\alpha}}{a_{0}^{\alpha}}\frac{\alpha^{2}%
}{(2-\alpha)^{4}}\sum_{i=1}^{3}\frac{\left[  1-\sqrt{1-4k_{i}^{2}\left(
\frac{2-\alpha}{\alpha}\right)  ^{2}}\right]  ^{2}}{k_{i}^{2}}. \label{f7}%
\end{equation}
 
In the thermodynamic limit, the total mean density $ \rho $ of this gas is given by 

\begin{eqnarray}
\rho=\lim_{V\rightarrow\infty}\frac{N}{V}
=\rho_{0}+\frac{1}{\piup^{3}a_{0}^{3}}%
\int_{0}^{+\infty}\mathrm{d}^{3}k\left[ \mathrm{e}^{\beta(E_{\alpha}-\mu)}-1\right]
^{-1}
=\rho_{0}+\rho_{e}\,, \label{l6}%
\end{eqnarray}
where $\rho_{0}$ is the occupation mean density in the ground state and $\rho_{e}$ is the occupation
mean density in the excited levels $\left(k\neq0\right)$ given by
\begin{equation}
\rho_{e}=\frac{1}{\piup^{3}a_{0}^{3}}\int_{0}^{+\infty} \mathrm{d}^{3}k\left[  \mathrm{e}^{\beta
(E_{\alpha}-\mu)}-1\right]  ^{-1}. \label{l10}%
\end{equation}
%a remark, we can prove that the condensation may exhibit in 3D and 2D cases except 1D case throughout  the last formula (\ref{f7}), the calculation of $\rho_{1}$ (the mean density of
%the excited bosons) from formula (\ref{l10}) for $T\leq T_{c}$ 
%

%
For $T=T_{c}$ ($ T_{c} $ is the critical temperature of condensation at which the chemical potential $\mu=0$), the total mean density of these bosons in the box ($V=L^{3}$)
becomes 

\begin{eqnarray}\label{111222333}
\rho&=&\frac{1}{\piup^{3} a_{0}^{3}}\int_{0}^{B}\int_{0}^{B}\int_{0}^{B} \frac{\mathrm{d} k_{1} \mathrm{d} k_{2} \mathrm{d} k_{3}}{\exp\left\lbrace \beta_{c} \frac{\hbar^{\alpha}D_{\alpha}}{a_{0}^{\alpha}} \frac{\alpha^{2}}{(2-\alpha)^{4}} \sum_{i=1}^{3} \frac{\left[ 1-\sqrt{1-4 k_{i}^{2}\left(\frac{2-\alpha}{\alpha} \right)^{2}}\right] ^{2}}{k_{i}^{2}} \right\rbrace -1}\nonumber\\
&=&\left[ \frac{\alpha}{2 \piup a_{0}(2-\alpha)}\right] ^{3}\int_{0}^{1}\int_{0}^{1}\int_{0}^{1}\frac{\mathrm{d} K_{1} \mathrm{d} K_{2} \mathrm{d} K_{3}}{\exp\left[\frac{4}{(2-\alpha)^{2}}\frac{\hbar^{\alpha}D_{\alpha}}{a_{0}^{\alpha}k_\text{B}}\frac{1}{T_{c}}\sum_{i=1}^{3}\left( \frac{1-\sqrt{1-K_{i}^{2}}}{K_{i}}\right)^{2}  \right] -1}\nonumber\\
&=&\left[ \frac{\alpha}{2 \piup a_{0}(2-\alpha)}\right] ^{3}\int_{0}^{1}\int_{0}^{1}\int_{0}^{1}\frac{\mathrm{d} K_{1} \mathrm{d} K_{2} \mathrm{d} K_{3} }{\exp\left[\frac{4\piup^{-\alpha}}{(2-\alpha)^{2}}\frac{E_{0}}{k_\text{B}}\frac{1}{T_{c}}\sum_{i=1}^{3}\left( \frac{1-\sqrt{1-K_{i}^{2}}}{K_{i}}\right)^{2} \right]-1},
\end{eqnarray}
where $K_{i}=2\left(2-\alpha\right)k_{i}/\alpha$ and $ D_{\alpha} $ is given as \cite{ali2}
 \begin{equation}
 D_{\alpha}=\frac{E_{0}}{\hbar^{\alpha}k_{0}^{\alpha}},\label{d}%
 \end{equation}
  where $E_{0}=\hbar^{2}k_{0}^{2}/2m$, $k_{0}=\piup/a_{0}$, and $ a_{0} $ are the characteristic energy, wavenumber and the length of the system, respectively, ($ a_{0}$ can be shown as the diameter of atoms). For $ \alpha=2$, the equation (\ref{d}) yields $ D_{2}=1/2m $. 
 From the above formula (\ref{111222333}), we plot the curves of $T_{c}$ against the density $\rho$ for various $ \alpha $ values as presented in figure \ref{fig.1}.  We choose $a_{0}=2.56\cdot10^{-10}~\mathrm{m} $ and $ m=6.64\cdot10^{-27}~\mathrm{kg} $ being the diameter and the mass of the helium atoms, respectively. In this case, $ k_{0}\approx 1.23 \cdot 10^{10}~\mathrm{m}^{-1} $ and $ E_{0}/k_\text{B}\approx 9.12$~K.

\section{Bose-Einstein condensation based on Laskin approach}
\label{s3}
In his paper \cite{laskin2000}, Laskin showed that bosons have the single-particle energy spectrum defined by
\begin{eqnarray}
 E_{\alpha}=D_{\alpha}\left(\piup\hbar\right)^{\alpha}\left[  \frac{n_{1}^{2}%
}{L^{2}}+\frac{n_{2}^{2}}{L^{2}}+\frac{n_{3}^{2}}{L^{2}}\right]
^{\alpha/2}
=D_{\alpha}\hbar^{^{\alpha}}k^{^{\alpha}}, \quad (n_{1}, n_{2}, n_{3}) \in \mathbb{Z} 
\label{l4}%
\end{eqnarray}
Subsequently, \cite{z14}  derived the critical temperature for an ideal Bose gas confined in a hard-wall box with the volume $V=[0,L]^{3}$ 
\begin{eqnarray}
 T_{c}=\frac{\hbar^{\alpha}D_{\alpha}}{k_\text{B}}\left[  \frac{2\piup^{2}\alpha\rho
}{\Gamma\left(  \frac{3}{\alpha}\right)\zeta\left(
	\frac{3}{\alpha}\right)  }\right]^{\alpha/3}
=\frac{E_{0}}{k_\text{B}k_{0}^{\alpha}}\left[\frac{2\piup^{2}\alpha\rho
}{\Gamma\left(  \frac{3}{\alpha}\right)\zeta\left(
	\frac{3}{\alpha}\right)}\right]^{\alpha/3}. \label{aaa} 
\end{eqnarray}
 
If we put $\alpha=2$, we obtain%
\begin{equation}
T_{c}=\frac{h^{2}}{2\piup mk_\text{B}}\left[\frac{\rho}{\zeta\left(\frac{3}%
	{2}\right)}\right]^{2/3}, \label{l12}%
\end{equation}
which is the well-known standard critical temperature since, in this case, the Laskin spectrum coincides with the standard one \cite{hang}.

\begin{figure}[!h]
	\begin{center}
		\includegraphics[width=9cm]{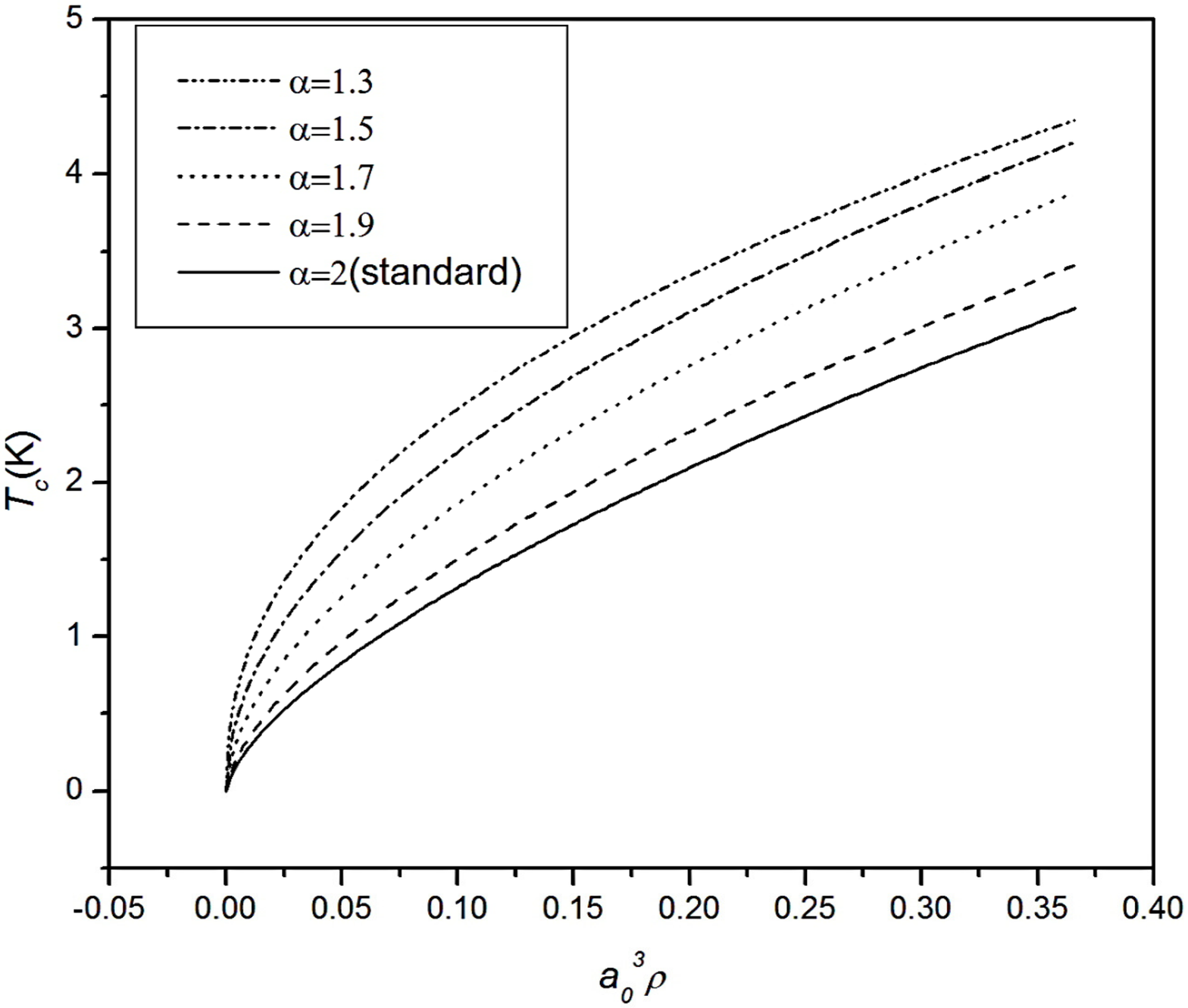}
		\caption{The critical temperature curves at different $\alpha$ values according to Laskin approach.}
		\label{fig.2}
	\end{center}
\end{figure}

We  plotted $ T_{c} $ given by equation~(\ref{aaa}) with respect to the density $ \rho $  [see figure \ref{fig.2}]  for some specific helium parameters ($ a_{0} $ and $ m $).

\section{Discussion}
\label{s4}
From figures \ref{fig.1} and \ref{fig.2}, we can deduce that: 
(i) The critical temperature $T_{c}$ grows with the density $\rho$ both in  non-local and in standard approaches as expected. 
(ii) In the non-local approach,  $T_{c}$ is increased for $ \alpha<2 $. 
It tends to the standard critical temperature $T_{c}$ for the case $ \alpha=2 $ where the Caputo-Fabrizio derivative is in complete agreement with the results known from the standard derivative. 
It turns out that the transition temperature obtained via the Caputo-Fabrizio approach is larger than the standard transition temperature and than the one obtained by using Laskin approach. For example when $ \alpha=1.3 $ and $ \rho\approx10^{28}~{\rm m^{-3}} $, $T_{c}$ exceeds  the standard one 10 times. 
In our opinion, the difference of critical condensation temperatures is due to the allowed number of energy levels in non-local case. 
In other words, the finite number of energy levels increases $T_{c}$. 
From the condition (\ref{f41}), this number decreases when $ \alpha $ decreases. Therefore, $T_{c}$ increases as well. 
On the contrary, the number of energy levels is infinite in the standard approach.

 \section{Conclusion}
\label{con}

The study of Bose-Einstein condensation from Caputo-Fabrizio point of view on the momentum power law has revealed various effects of the non-local derivative on the critical condensation temperature. 
This goal is reached by considering an ideal bosonic gas enclosed in a hard-wall box for a different energy spectrum. 
This spectrum is derived based on a non-local Schr\"{o}dinger equation defined in terms of Caputo-Fabrizio derivative.
The main findings can be summarized as follows: (i)  The number of energy levels is bounded and related to the derivative order $ \alpha $.
(ii) As a result of  the bounded number of energy levels, the critical temperature for $ \alpha<2 $ is greater than the standard one. 
(iv) When $\alpha=2$, all the results become standard because, in this case, the Caputo-Fabrizio derivative collapses to the ordinary one.
(v) Additionally, the condensation for $ \alpha<2 $  is possible in two dimensions.
The last finding will be investigated more in  detail in our next research paper where the thermodynamic properties of such systems will be reviewed.  
 Furthermore, new effects might be highlighted by extending this study to other quantum statistical systems such as Fermi gas.

 To compare, we have presented the transition temperature based on the Laskin spectrum of the ideal Bose gas. It turns out that this temperature is grater than the standard temperature transition but smaller than those obtained from the Caputo-fabrizio spectrum.

\section*{Acknowledgements}
All authors would like to thank the head of the LABTHOP laboratory, Professor Mansour Abdelouahab, Mathematics Department at University of El-Oued, Algeria, for
the material support.

%\subsection{References}

\newpage
\ukrainianpart

\title{Вплив нелокальної похідної на конденсацію Бозе-Ейнштейна%
}
\author{Ф.Е. Бузенна\refaddr{label1},
        M.T. Мефтах\refaddr{label2}, M. Діфалах\refaddr{label3}}
\addresses{ 
\addr{label1} Фізичне відділення і лабораторія LEVRES, факультет точних наук, університет  Ель Уеда, 39000, Алжир\\
%e-mail: fatmahp2000@gmail.com
\addr{label2} Фізичне відділення і лабораторія LEVRES, факультет математики і матеріалознавства, університет Касді Мербах, Уаргла 30000, Алжир\\
%e-mail: mewalid@yahoo.com
\addr{label3}Фізичне відділення і лабораторія LEVRES, факультет точних наук, університет  Ель Уеда, 39000, Алжир\\
%e-mail: mosdifah@yahoo.fr
}
\makeukrtitle 
\begin{abstract}
У цій роботі ми вивчаємо вплив нелокальної похідної на конденсацію Бозе-Ейнштейна.
Спершу, ми розглядаємо похідну Капуто-Фабріціо дробового порядку $ \alpha $ для виведення власних значень нелокального рівняння Шредінгера для вільної частинки в 3D боксі.
Потім ми розглянемо 3D-конденсацію Бозе-Ейнштейна ідеального газу з отриманим енергетичним спектром.
Цікаво, що при такому підході критичні температури $T_ {c}$ конденсації для $ 1 <\alpha <2 $ перевищують стандартну.
Крім того, показано, що конденсація в 2D є можливою.
По-друге і для порівняння, ми представили на основі спектру, встановленого Н.~Ласкіним, критичну температуру переходу як функцію дробового параметра $\alpha$ для системи вільних бозонів, керованих гамільтоніаном із степеневим законом щодо моменту ($H\sim p ^{\alpha} $). У цьому випадку ми продемонстрували, що температура переходу перевищує стандартну.
Порівнюючи дві  температури переходу (по відношенню до Капуто-Фабріціо та Ласкіна), ми виявили для фіксованих $\alpha $ та густини $\rho$, що температура переходу по відношенню до Капуто-Фабриціо вища, ніж по відношенню до Ласкіна.	
	

\end{abstract}

\end{document}